\documentstyle[aps,epsfig,multicol]{revtex}

\begin{document}
\draft

\title{\bf Raman cooling and heating of two trapped Ba$^+$ ions}
\author{D. Rei\ss, K. Abich, W. Neuhauser, Ch. Wunderlich, P.E. Toschek}
\address{Institut f\"ur Laser-Physik, Universit\"at Hamburg, Jungiusstr. 9, 
20355 Hamburg, Germany}
\date{\today}
\maketitle
\begin{abstract}
  We study cooling of the collective vibrational motion of two
  $^{138}$Ba$^+$ ions confined in an electrodynamic trap and
  irradiated by
  laser light close to the resonances S$_{1/2}$-P$_{1/2}$ (493 nm) and
  P$_{1/2}$-D$_{3/2}$ (650 nm). The motional state of the ions is
  monitored by a spatially resolving photomultiplier. Depending on
  detuning and intensity of the cooling lasers, macroscopically
  different motional states corresponding to different ion
  temperatures are observed. We also derive the ions' temperature from
  detailed analytical calculations of laser cooling taking into
  account the Zeeman structure of the energy levels involved. The
  observed motional states perfectly match the calculated
  temperatures. Significant heating is observed in the vicinity of the
  dark resonances of the Zeeman-split S$_{1/2}$-D$_{3/2}$ Raman
  transitions. Here two-photon processes dominate the interaction
  between lasers and ions. Parameter regimes of laser light are
  identified that imply most efficient laser cooling.
\end{abstract}

\begin{multicols}{2}

\section{Introduction}
\label{sec:Intro}
In order to implement conditional quantum dynamics of qubits
represented by the internal structure of individual ions in a Paul
trap, a vibrational mode of the entire ion 
string 
is used. Cirac and Zoller \cite{Cir95} have suggested a method for
conditional dynamics that requires the ions to be cooled to their
vibrational ground state. Here, the vibrational state represents a
``bus'' qubit that gets entangled with the ions' internal states by
stimulated transitions, where the absorption (or emission) of a photon
is coupled with the emission (or absorption) of a vibrational quantum.

The recently proposed S{\o}rensen-M{\o}lmer scheme \cite{Sor99}
incorporates two-photon transitions in order to directly entangle the
internal states of the ions. Here, the vibrational mode is only
virtually excited. In this way the contribution of higher vibrational
excitations can be made to destructively interfere. In this way, the
scheme does not require cooling to the vibrational ground state, but
admits population in higher vibrational levels.

In general,
the
manipulation of the motion of trapped particles by light relies on the
coupling of electronic and motional degrees of freedom via the
scattering of photons. Here, the important figure of merit is the
Lamb-Dicke parameter (LDP) $\eta$, its square given by the ratio of
the recoil energy transfered when scattering a photon, and the energy
of the vibrational quantum \cite{Ste86}. For small $\eta$, transition
processes involving the creation or annihilation of $n$ vibrational
quanta (i.e. $n$th order sideband transitions) are
suppressed by the factor $\eta^{2n}$. Therefore, a small LDP allows
only linear coupling (involving only single-phonon transitions)
between electronic and vibrational degrees of freedom. This
Jaynes-Cummings type of coupling \cite{Sho93} is prerequisite for many
schemes suggested in quantum information processing  including the
mentioned implementations \cite{Cir95,Sor99}.

Plenio and Knight \cite{Jon00} have proposed to drive the interaction
between electronic and vibrational degrees of freedom by tuning  
{\em strong} laser light to the resonance
frequency,
instead of the sideband. In this case, the light intensity imposes a
specific light shift on the ions that selects the desired
transition. By doing so, the time needed for conditional dynamics is 
reduced, as compared with the Cirac-Zoller scheme, thus reducing the
effect of decoherence. Here, too, the LDP plays a crucial role as the
fidelity of such a fast quantum gate is $f\approx 1-\case12\eta^2$.

While the S{\o}rensen-M{\o}lmer scheme does not require ground-state cooling,
it still demands low vibrational quantum numbers
(\mbox{i.e. $\eta^2(n+1)\ll1$}). 
Thus, laser cooling is
essential
for quantum information processing in ion traps, and it is desirable
to have at hand robust cooling techniques that are suited
for reaching
low vibrational excitation at moderate technological expenditure on
the side of the laser sources. In particular, cooling of the
collective motion of several particles is prerequisite for
implementing conditional quantum dynamics with trapped ions.

It has been known for some time that 
the motion of trapped particles strongly depends on 
laser cooling.
With
a single trapped particle, distinct macroscopic motional states have
been observed with different parameters of the cooling lasers \cite{Sau88}.
With
a cloud of ions, changes in the temperature induced by laser
cooling cause phase transitions between a crystalline and an
evaporated state \cite{Blu88,SaG88}. These macroscopically distinct
states indicate the specific temperature achieved by laser cooling.

In this paper we investigate the macroscopic motional states of two
trapped $^{138}$Ba$^+$ under the action of laser cooling. In
Sec.~\ref{sec:Theory}, the theory of laser cooling of the collective
motion of two charged particles trapped in  an anisotropic harmonic
potential is outlined. If the particles obey the Lamb-Dicke limit, the
cooling rate and the achievable temperature are calculable without any
free parameter. We fully take into account the multilevel structure of
the involved ions, thus allowing for any type of cooling process, such
as Doppler \cite{Neu78}, sideband\cite{Mon95}  and Raman cooling
\cite{Gil88}. In Sec.~\ref{sec:Exp} we describe the experimental setup
and present the results of the measurements. Phase transitions between
macroscopically distinct states are observed. The parameter values at
which these transitions take place perfectly agree with the values
calculated from the theoretical model. In Sec.~\ref{sec:Optimize} we
use the model to find optimized laser parameters for cooling the
collective motion of the ions.

\section{Theory}
\label{sec:Theory}

\subsection{Collective Motion}
\label{subsec:motion}
We consider the collective motion of two ions of mass $m$ in a
three-dimensional (3D)
harmonic trapping potential 
\begin{equation}
  V_{\text{trap}}({\bf r}_j)
  =\case m2 \left( \omega^2_x x_j^2+ \omega^2_y
    y_j^2+ \omega^2_z z_j^2\right),
\end{equation}
which is the same for both ions ($j=1,2$). Since 
each of the ions carries
one positive elementary charge $e$, they repel each other by the
Coulomb interaction 
\begin{equation}
  V_{\text{C}}({\bf r}_1,{\bf r}_2)=\frac{e^2}{4\pi\epsilon_0}
  \frac1{|{\bf r}_1-{\bf r}_2|}\quad.
\end{equation}
The potential is simplified by using center-of-mass and relative
coordinates 
\begin{equation}
  {\bf  R}\equiv\frac{{\bf  r}_1+{\bf  r}_2}2
\quad\mbox{and}\quad{\bf r}\equiv{\bf r}_2-{\bf r}_1\,.
\label{eq:cm_rm}
\end{equation}
With the definition of total and reduced masses $M=2m$ and $\mu=m/2$,
one gets
\begin{equation}
  V({\bf  R})=\case M2\left(\omega^2_x X^2+ \omega^2_y
    Y^2+ \omega^2_z Z^2\right)
\end{equation}
for the center-of-mass motion, corresponding to the motion of a single
trapped particle with the original oscillation frequencies, but twice
the ions mass.

The potential for the relative motion is 
\begin{equation}
  V({\bf r})=  \case\mu2\left(\omega^2_x x^2+\omega^2_y 
    y^2+\omega^2_z z^2\right)\;+\;
  \frac{e^2}{4\pi\epsilon_0}\frac1{|{\bf r}|}\,.
\end{equation}
If the trap frequencies slightly differ (as they do in our experiment,
see Table \ref{tab:LDP}) this potential does not show spherical
symmetry. With $\omega_x<\omega_y<\omega_z$ holding, the potential has
absolute minima on the $x$ axis and saddle points on the $y$ and
$z$ axis at 
\begin{equation}
  \label{eq:r_0}
  {\bf r}_{0,q}
  =\pm\sqrt[3]{\frac{e^2}{4\pi\epsilon_0\;\mu \omega_q^2}}\;{\bf e}_q
  \quad,\quad q=x,y,z\;.
\end{equation}

\begin{figure}[htbp]
  \begin{center}
    \epsffile{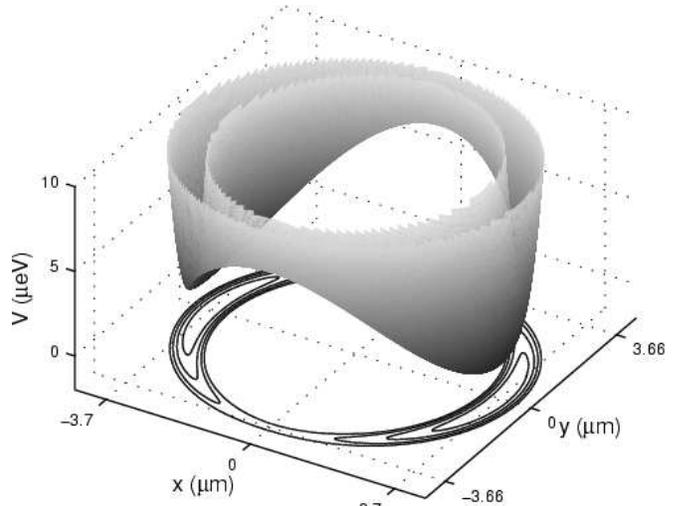}
  \caption{Effective double well potential $V(x,y)$ for the relative
    motion of two trapped $^{138}$Ba$^+$ ions (trap frequencies as
    given in Table \ref{tab:LDP}). The contour lines
    correspond to $V=2,4,6,8 \;\mu$eV.} 
  \label{fig:sattelpott}
  \end{center}
\end{figure}

Figure \ref{fig:sattelpott} shows the potential in the $(x,y)$ plane. It is
clear that for high enough temperature, i.e. at a mean motional energy
$  E_{\text{ex}}$ comparable with the potential barrier at the saddle
point on the $y$-axis 
\begin{equation}
  \label{eq:Pot_barrier}
  \Delta V_y=
  \frac32\sqrt[3]{\left(\frac{e^2}{4\pi\epsilon_0}\right)^2\mu}\;
  \left(\omega_y^{2/3}-\omega_x^{2/3}\right)\;,
\end{equation}
the ions can move freely in a ring-shaped region in the
$(x,y)$ plane. 
However, at low temperature, the relative motion is
restricted to the wells, i.e., the ions will appear at approximately
fixed positions. The same argument can be applied to the
$z$-direction: For even higher energies $E_{\text{ex}}\gtrsim\Delta
V_z$ the motional state changes from a ring to a sphere. All three
motional states have been observed in our experiment. However, in the
following we shall focus on the transition between the localized and
the ring state, as we are mainly interested in laser cooling in the
low-temperature regime.

In the localized state the potential can be approximated around the
minimum at ${\bf r}_{0,x}$. Expanding up to second order in the
displacement  $\tilde{\bf r}\equiv {\bf r}-{\bf r}_{0,x}$ one gets 
\begin{equation}
  V({\bf r})=V({\bf r}_{0,x})+\case\mu2\left(\tilde\omega_x^2 \tilde x^2
    +\tilde\omega_y^2 \tilde y^2+\tilde\omega_z^2\tilde z^2\right)
\end{equation}
with local frequencies
\begin{equation}
  \tilde\omega_x=\sqrt{3}\omega_x\quad,\quad
  \tilde\omega_{y,z}=\sqrt{\omega_{y,z}^2-\omega_x^2}
\end{equation}
(c.f. Table \ref{tab:LDP})
\begin{table}
 \begin{tabular}{l|cccccc}
Mode &$X$& $Y$& $Z$&$\tilde x$&$\tilde y$&$\tilde z$\\
\hline
$\omega/2\pi$MHz&1.0035&1.0220&1.0530&1.7381&0.1936&0.3191\\
$\eta_{493}$&0.044&0.044&0.043& 0.034&0.101&0.079\\
$\eta_{650}$&0.034&0.033&0.033& 0.026&0.077&0.060\\
$\eta_{1762}$&0.012&0.012&0.012& 0.009&0.028&0.022\\
 \end{tabular}
\caption{Frequencies and Lamb-Dicke parameters for the different
  vibrational modes. Note that the geometrical factor 
$({\bf k}_l\cdot {\bf e}_q)$ 
is included in the LDPs (Eq.~\ref{eq:Def_LDP}). In the present
experiment the angle of incidence is 45$^\circ$ to all directions,
yielding a factor of $1/\sqrt3$.}
\label{tab:LDP}
 \end{table}

\subsection{Laser Cooling of the collective motion}
\label{subsec:Cooling}

The mechanical effect of irradiating ions with laser light leads 
to thermalization of the motional degrees of freedom. The 
corresponding cooling rate and steady-state temperature depend on 
the electronic level scheme and laser parameters; they are 
calculated in this section. The key to the theory presented here 
is the well-known Lamb-Dicke limit: As long as the ions are 
localized near the minima of the potential, their spatial 
extension is much smaller than the wavelength of the exciting 
laser light. In this case, the photon recoil operator 
$\text{e}^{i{\bf k}_l\cdot{\bf r}}$, i.e. the coupling of the the 
internal dynamics to the ions motion, can be expanded in the LDP. 
Since the LDP is small, dynamics 
due to 
this coupling is slow compared with the internal dynamics that may be
adiabatically eliminated. This reduction leads to a master equation
for the motional degrees of freedom, where the involved transition
rates depend on steady-state expectation values of internal operators.
This procedure and its application to a single Ba$^+$ ion has been
described in \cite{Rei96}.

We do not account for dipole-dipole interaction between the two 
ions, since the separation of the two ions is large as compared to 
the optical wavelengths used ($r_0/\lambda\approx8$.) Even though 
processes caused by this interaction
possibly could be observed \cite{Bei99},
they are rare, compared with the single-ion scattering, and their
influence on the collective motion is negligible. Thus, we do not
consider correlations between the two ions' internal dynamics that
would lead to cooperative effects such as superradiance. We rather
calculate the cooling effect that each of the two ions individually
has on the common motion. This effect is the same for both ions, since
the foci of the lasers are wide compared with the distance of the ions
(the waist of the Gaussian beams being $w \approx 10 \,r_0$), and
therefore the laser intensities, and consequently laser cooling do
not depend on the ions' positions.

We shall now give a short recap of the theory presented in 
Ref.~\cite{Rei96} and apply its results to the collective motion of two
ions in the current experiment. We start from the density operator
$\varrho$ for the motional and internal degrees of freedom. The
dynamics is described by the master equation 
\begin{equation}
  \label{eq:Master}
  \frac d{dt}\varrho(t)=
  \case1{i\hbar}[H,\varrho(t)]+
  {\cal L}\varrho(t)\quad ,
\end{equation}
where 
\begin{eqnarray}
  {\cal L}\varrho&=&\sum_{j=1,2}
  \sum_l \gamma_l\bigg(2\sigma^-_{l,j}
  \int_{-1}^{1}\!\! d\zeta W_l(\zeta)
  \text{e}^{-i k_l r_j\zeta}
  \varrho
  \text{e}^{i k_l r_j\zeta}\sigma^+_{l,j}
  \nonumber\\
  &&\hphantom{\sum_{j=1,2}
    \sum_l \gamma_l\bigg(}
  -\;\sigma^+_{l,j}\sigma^-_{l,j}\,\varrho
  \;-\;\varrho\,\sigma^+_{l,j}\sigma^-_{l,j}\bigg)
\end{eqnarray}
is the Liouvillean allowing for spontaneous emission. It has the 
usual Lindblad form \cite{Lin76}, where dissipation acts on the 
internal (via $\sigma^\pm$) and on the motional (via 
$\text{e}^{\pm i{\bf k\cdot r}}$) degrees of freedom. The index 
$j$ labels the two ions, and the index $l$ runs over all atomic 
transitions involved. $2\gamma_l$ is the spontaneous emission 
rate, and $\sigma^\pm_l$ are the corresponding transition 
operators. $W_l(\zeta)$ is the angular distribution of 
spontaneous emission for this transition. The Hamiltonian is 
given by
\begin{equation}
  H=\frac{{\bf P}^2}{2M}+V({\bf R})+\frac{{\bf p}^2}{2\mu}+V({\bf r})
+\sum_{j=1,2}H_{\text{in},j}+V_{\text{Dip},j}\,.
\end{equation}
$H_{{\rm in},j}$ is the Hamiltonian for the internal levels 
of the $j$-th ion. 
The ion's dipole interaction with the laser
light can be written as
\begin{equation}
V_{\text{ Dip},j} = \sum_l \hbar\Omega_l 
\;\text{e}^
{i {\bf k}_l\cdot{\bf r}_j}\;\sigma^+_{l,j}\; +\;\mbox{H.c.}\;,
\end{equation}
where $\Omega_l$ is the Rabi frequency of the light driving the
transition $l$.  
Summing over both ions, and rewriting the dipole interaction in terms
of center-of-mass and 
relative coordinates [(Eq.~\ref{eq:cm_rm})], gives
\begin{equation}
V_{\text{ Dip}} = \sum_l \hbar\Omega_l \;\text{e}^{i{\bf k}_l
  \cdot{\bf R}}
\left(\text{ e}^{-\frac i2{\bf k}_l\cdot{\bf r}}\sigma^+_{l,1}
+\text{e}^{\frac i2{\bf k}_l\cdot{\bf r}}\sigma^+_{l,2}
\right)
 +\mbox{H.c.}
\end{equation}
Note that the center-of-mass
and relative motions are coupled via the interaction with the laser light. 
However, in the Lamb-Dicke limit this
coupling can be neglected, as 
$\langle {\bf k}_l\cdot{\bf R}\rangle$, and 
$\langle{\bf k}_l\cdot{\bf \tilde r}\rangle$ are small and the
potential can be expanded. This expansion is characterized by the
Lamb-Dicke parameters
\begin{equation}
\label{eq:Def_LDP}
  \eta_{l,q}={\bf k}_l\cdot{\bf e}_q\sqrt{\hbar/2m\omega_q}
\end{equation}
for the center-of-mass modes
and $ \tilde\eta_{l,q}={\bf k}_l\cdot{\bf e}_q
\sqrt{\hbar/2m\tilde\omega_q}$
for the normal modes of the relative motion. 
In the experiment, all LDPs are much smaller than unity, 
see Table \ref{tab:LDP}. Now, we rewrite the coordinates in
terms of the lowering and raising operators of the
corresponding vibrational eigenmodes 
\begin{equation}
 R_q=\sqrt{\frac{\hbar}{2 M \omega_q}}\,(a_q^\dagger+a_q)\;,\;
 \tilde r_q=\sqrt{\frac{\hbar}{2 \mu
     \tilde\omega_q}}\,(\tilde a_q^\dagger+\tilde a_q).
\end{equation}
Since the expansion is around the potential minimum, there is a 
constant phase factor 
$\text{ e}^{\pm\frac i2{\bf k}_l\cdot{\bf r}_0}$, that is
incorporated in the atomic transition operators $\sigma^+_{l,j}$.
Now, keeping only terms up to first order in (any) Lamb-Dicke parameter
yields
\begin{eqnarray}
  V_{\text{dip}}&=& \sum_{j=1,2}\sum_l 
 \hbar\Omega_l\sigma^x_{l,j}
-\frac1{\sqrt2}
\sum_q\hbar\Omega_l\eta_{l,q}(a_q^\dagger+a_q)\;\sigma^y_{l,j}
\nonumber\\
&&-\frac{(-1)^j}{\sqrt2}
\sum_q\hbar\Omega_l\tilde\eta_{l,q}(\tilde a_q^\dagger+\tilde a_q)
\sigma^y_{l,j}\,,
\label{eq:V_dip}
\end{eqnarray}
where $\sigma^x=\sigma^++\sigma^-$ and
$\sigma^y=(\sigma^+-\sigma^-)/i$ have been used. 
The first term of Eq.~(\ref{eq:V_dip}) 
describes the driving of the
internal dynamics by the interaction with the laser light. Since it is of
zeroth order in any LDP it does not contain the mechanical effect of light,
i.e. no laser cooling or heating. 
The second and third term of Eq.~(\ref{eq:V_dip}) 
(first order in a LDP) couple the internal
dynamics to the center-of-mass and relative motions, respectively.
The ion-dependent sign in the third line of Eq.~(\ref{eq:V_dip}) arises
from the opposite impact of the two ions' recoil on the relative motion.
Since the vibrational modes decouple in the Lamb-Dicke limit, we can
restrict ourselves to a 1D description. Therefore, the
index $q$ characterizing the specific mode is omitted in what follows.

Now, the fast internal dynamics of the ions is adiabatically
eliminated 
and one obtains a master equation for the external degrees
of freedom. 
This procedure finally leads to the well known 
time evolution of the mean phonon number \cite{Lin86}
\begin{equation}
  \frac{d}{dt}\langle n\rangle=-(A_--A_+)\langle n\rangle+A_+\quad ,
\end{equation}
with the cooling rate
\begin{equation}
  W=A_--A_+
\end{equation}
and (if the cooling rate is positive) the mean motional energy 
\begin{equation}
  E_{\text{Ex}}=\hbar \omega\frac{A_+}{A_--A_+}\quad .
\end{equation}
The transition rates $A_\pm$ are 
given by
\begin{equation}
  A_\pm=2 (\text{Re}S(\mp\omega)+D)\quad .
\end{equation}
The first constituent is
the fluctuation spectrum of the electronic
dipole coupling
\begin{equation}
  S(\omega)= \sum_{l,l'}\Omega_l\Omega_{l'}\eta_l\eta_{l'}
  \int_0^\infty\!\!\!\! dt\text{e}^{i\omega t}
\langle\sigma^y_{l}(t)\sigma^y_{l'}(0)\rangle\,,
\end{equation}
where $\omega$ is the vibrational frequency of the considered mode.
The expectation value is to be taken in the steady state  of the
internal dynamics that in turn is calculated from the master
equation 
\begin{equation}
  \frac d{dt}\varrho_{\text{in}}(t)=
\frac1{i\hbar}\left[H_{\text{in}}+\sum_l\hbar \Omega_l
  \sigma^x_l\,,\,\varrho_{\text{in}}(t)\right]+ 
{\cal L}_0\varrho_{\text{in}}(t)\quad ,
\end{equation}
derived from Eq.~(\ref{eq:Master}), by using adiabatic 
elimination, and keeping only
the zeroth order in any LDP.
Here,
\begin{equation}
{\cal   L}_0\varrho_{\text{in}}=
\sum_l \gamma_l(2\sigma^-_l\varrho_{\text{in}}\sigma^+_l
-\sigma^+_l\sigma^-_l\varrho_{\text{in}}
-\varrho_{\text{in}}\sigma^+_l\sigma^-_l)
\end{equation}
describes the spontaneous emission in zeroth order.
Via the driving by laser light, the steady state depends on the
parameters (intensity, detuning, polarization) of the light field. 

The second contribution to the transition rates originates from the
spontaneously emitted photons as described by the
Liouvillean ${\cal L}$. So far, spontaneous emission has only
been considered up to the zeroth order in the LDP. The second order term
is readily calculated and leads to the additional transition rate 
\begin{equation}
  D=\sum_l\gamma_l\,\frac{\alpha_l}{({\bf k}_l\cdot {\bf e}_q)^2}\,\eta_l^2\, 
\langle\sigma^+_l \sigma^-_l\rangle
\end{equation}
both from $|n\rangle$ to $|n+1\rangle$ and vice versa.
Here, the geometrical factor $({\bf k}_l\cdot {\bf e}_q)^2$
in the LDP (eq. 15) is replaced by
$\alpha_l=\int_{-1}^1d\zeta W_l(\zeta)\zeta^2$ that accounts for the
angular distribution of the light emission.

\section{Experimental Observation}
\label{sec:Exp}
The two Ba$^+$-ions are confined in a 1-mm-diameter Paul trap and
irradiated by laser light at 493 nm for excitation of resonance
scattering on the $^2$S$_{1/2}$-$^2$P$_{1/2}$ transition. This laser
is detuned slightly below resonance for cooling the ions. The
fluorescence signal is recorded by photon counting. A second laser at
650 nm prevents optical pumping into the $^2$D$_{3/2}$ level. A static
magnetic field defines the quantization direction and lifts the
degeneracy of the magnetic sublevels. Its direction is at 45 degrees
with respect to the plane defined by the ring electrode. The
directions of propagation and the polarizations of both light beams
are set perpendicular to the magnetic field. The power levels of the
light fields are stabilized by electro-optic modulators. Precise
detuning of the 650-nm laser is accomplished using a double-pass
acousto-optic modulator. So far, the setup has been described
elsewhere \cite{Sch95}. In addition the fluorescent light is
monitored by a spatially resolving photo multiplier revealing the
position of the ions in the trap.

The ions are dynamically trapped by superimposed electric dc and
rf-fields \cite{Neu80} and represent particles inside a nearly spheroidal
harmonic (quasi)potential. In order to determine the geometry of this
(quasi)potential oscillations of a single trapped ion are excited by
applying an additional rf field. Three sharp resonances are found
at the frequencies given in Table \ref{tab:LDP},
i.e. macroscopic motion of the ion in the $x$, $y$ and/or $z$
direction of the trap is observed. The $x$-direction,
within the radial plane, is found at approximately 
45 degrees to the polarization of the lasers.
While the nondegeneracy of $\omega_z$ is obvious, the asymmetry
of $\omega_x$ and $\omega_y$ originates from a slight asymmetry
of the ring electrode.
The distance of the two ions in the potential minima, as calculated
from Eq.~(\ref{eq:r_0}) using the actual values of $\mu$
and $\omega_q$ is $r_0=3.70\,\mu$m and agrees
very well with the experimental value $3.65\,\mu$m obtained
from the spatially resolved measurement.
\begin{figure}[htbp]
  \begin{center}
\epsffile{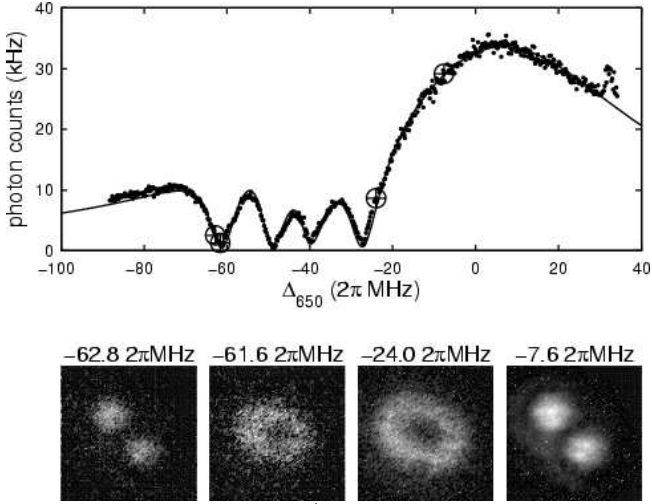}
    \caption{Two trapped Ba$^+$ ions show different motional states
      depending on laser parameters. Top: fluorescence of two
      trapped ions as a function of laser detuning, collected in
      0.1s. Bottom:
      spatial distribution of the two ions at the detunings indicated
      above.}
    \label{fig:anreg_mcas}
  \end{center}
\end{figure}
Keeping  all other parameters constant, the frequency of the 650-nm laser
is scanned across the $^2$D$_{3/2}$-$^2$P$_{1/2}$ line. 
The intensity of the scattered light vs detuning 
of the 650-nm light, 
$\Delta_{650}=\omega_{650}-\omega_{\text{P}\leftrightarrow\text{D}}$,
is shown in the upper part of 
Fig.~\ref{fig:anreg_mcas},
where $\omega_{650}$ is the laser frequency and
$\omega_{\text{P}\leftrightarrow\text{D}}$ is the atomic transition
frequency. The structure of dark resonances is well
known from the fluorescence of a single ion \cite{Sta98} and 
corresponds to four dark states, i.e.~ coherent superpositions of
magnetic substates of the $^2$S$_{1/2}$ and $^2$D$_{3/2}$ levels,
respectively. From a fit of this excitation spectrum with eight-level
optical Bloch equations, accounting for the Zeeman substructure,  
we obtain the Rabi frequencies of both lasers, the detuning
$\Delta_{493}=\omega_{493}-\omega_{\text{S}\leftrightarrow\text{P}}$ 
of the 493 nm laser as well as the strength of the
magnetic field. This  
is a standard procedure in our Ba$^+$ experiments. The parameters
obtained from the spectrum shown in Fig.~\ref{fig:anreg_mcas} are
$\Omega_{650}=59.0\times2\pi$ MHz, 
$\Omega_{493}=46.2\times2\pi$ MHz,
$\Delta_{493}=-44.3\times2\pi$ MHz and
$B/\hbar\mu_{\text{B}}=7.7\times2\pi$ MHz, where 
$\mu_{\text{B}}$ is the Bohr magneton.

Additional information is derived from monitoring the motion
of the two ions. Depending on the detuning of the 650-nm laser,
different motional states are observed. Whereas the ions appear at
fixed positions if the red laser is tuned below the lowest or well above
the highest dark resonance, they form a
ring-shaped distribution in the $(x,y)$ plane with certain 
intermediate detunings. 

The lower part of Fig.~\ref{fig:anreg_mcas} shows the motional state
at four different detunings, corresponding to the crosshairs in
the spectrum. Note the clear distinction of the localized state at
$\Delta_{650}=-62.8\times2\pi$ MHz from the non-localized state only
$2.6\times2\pi$ MHz away: Phase transitions between the two observed
motional states appear at very definite laser frequencies. 
\begin{figure}[htbp]
  \begin{center}
\epsffile{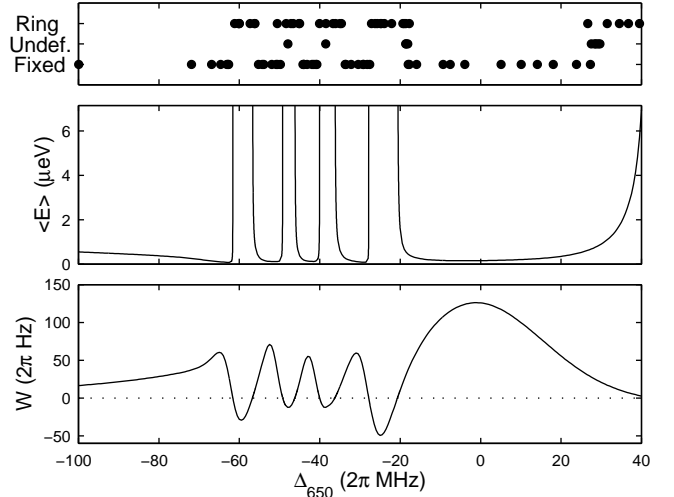}
    \caption{Top: observed motional states for different
      detunings of the 650-nm light. The dots correspond to
      individual observations. 
      Middle: mean motional energy in the $\tilde y$-mode
      calculated from theory.
      Bottom: cooling rate for the $\tilde y$-mode
      calculated from theory.}
    \label{fig:beob}
  \end{center}
\end{figure}

Obviously, the motional state of the ions depends on the ion temperature
set by the laser cooling. In Fig.~\ref{fig:beob}, the 
motional states observed at different detunings are shown together
with the motional energies and cooling rates calculated from theory.
The different motional states appear in distinct regions of detuning:
The ions are observed in the fixed position for the detuning
$\Delta_{650}$ below $-61\times2\pi$ MHz. At this limiting value
a sharp phase transition
occurs, and the ions form a ring when 
$ -61\times2\pi$ MHz $<\Delta_{650}< -56\times2\pi$ MHz. 
At the upper boundary, they are cooled again to the fixed-position state.
Such a ``hot'' region appears with each dark resonance.
At some values of the detuning, e.g.~at 
$\Delta_{650}=-19\times2\pi$ MHz,
the motion fluctuated between
the fixed and the ring-shaped state, thus remaining indefinite.

The motional energy has been calculated from the model presented in the
preceding section, using the experimental values for the $\tilde y$-mode. 
This energy should be compared with the height of the potential
barrier [(Eq.~\ref{eq:Pot_barrier})], which for the given data is
$\Delta V_y=7.15\;\mu$eV. Wherever the predicted temperature varies
steeply upon detuning, a sharp transition is observed. This is the
case, when the corresponding cooling rate changes from positive to
negative, i.e. when the lasers actually heat the ion. Laser heating
appears at {\em negative} detunings of both lasers when Raman
transitions dominate the interaction. In the vicinity of the dark
resonances, the internal state consists of a coherent superposition of
the S$_{1/2}$ and the D$_{3/2}$ states, forming an effective two level
system. Now, with both detunings being negative, and
$\Delta_{650}>\Delta_{493}+\Delta_{\text{zee}}$, the effective
detuning of this Raman transition is positive, and causes heating of
the motion. Here, $\Delta_{\text{zee}}$ is the difference in Zeeman
shifts of the involved sublevels that defines the resonance condition
of the considered dark state.

Note that the transition to the ring state, at positive detunings from
the Raman resonances, appears where the expected energy is still
considerable lower than the potential barrier. However, this does not
contradict the model, which assumes the ions close to the potential
minima. As soon as they are smeared out over a region of considerable
size, the harmonic approximation around the minimum no longer
holds. The cooling  gets less effective for the broadened velocity
distribution. This happens when the thermal energy is substantially
lower than $\Delta V_y$. Still, the observed phase transitions agree
very well with the results of the model. 

At low speed of scanning, for example, $d\Delta_{650}/dt\approx
1\times2\pi$ MHz/s, we observed hysteresis effects: 
The motional state upon up-tuning the laser frequency differed from
the state observed upon down-tuning.
However, this effect vanished when the scanning rate was decreased. Scans
at a very low speed 
($d\Delta_{650}/dt\lesssim 0.1\times2\pi$ MHz/s)
showed that the motional state 
solely depends on the laser parameters. 
Therefore, a steady state exists, which is approached in a finite
motional relaxation time. 

The final temperature is, in principle, not only set by the laser
cooling, but it is subject to additional heating mechanisms, such as
fluctuations in the trapping electric field. However, the calculated
transition rates $A_\pm \approx 10^5\ldots10^6$ s$^{-1}$ equal 
the number of scattered phonons
per unit time, and thus exceed the rate of parasitic 
heating ($\approx 10\ldots10^4$ s$^{-1}$) observed in
ion trap experiments \cite{Roh00}. Therefore, this additional
heating can be neglected as long as the ions are kept laser-cooled.

\section{Optimization of Laser Cooling}
\label{sec:Optimize}
\begin{figure}[htbp]
  \begin{center}
\epsffile{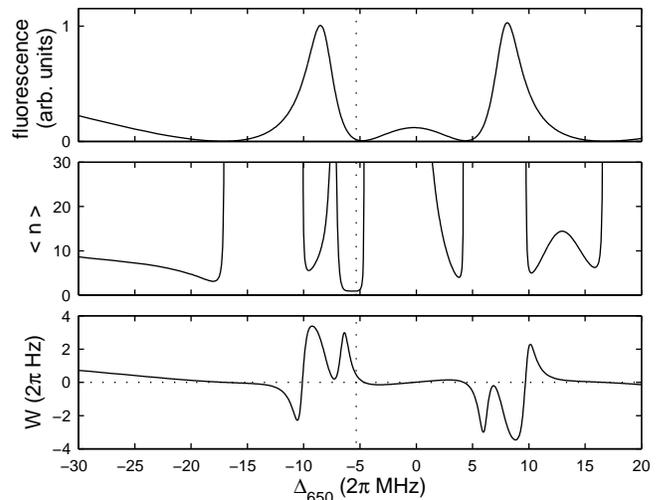}
    \caption{Fluorescence, mean number of vibrational quanta 
      in the $\tilde x$-mode, and
      cooling rate vs laser detuning. The vertical
      dotted line indicates the minimal $\langle n\rangle$,
laser parameters: $\Omega_{650}=89.6\times2\pi$ MHz, 
$\Omega_{493}=2.1\times2\pi$ MHz, $\Delta_{493}=-0.21\times2\pi$
MHz. Other parameters as in Sec.~\ref{sec:Exp}}  
    \label{fig:ApAmin1}
  \end{center}
\end{figure}

The discussed model is now employed to numerically investigate the
influence of various parameters on the laser cooling, and to find
optimized conditions for effective and robust cooling.
We concentrate on the $\tilde x$ \mbox{(stretch)mode}, since  
it is less affected by perturbations from stray fields of
the trap \cite{Kin98} and therefore is  suitable for serving as a bus
mode  between the ions.
A promising candidate for implementing 
qubits in trapped Ba$^+$ is the S$_{1/2}$-D$_{5/2}$ (1762 nm)
transition \cite{Apa98}. The LDP corresponding to this infrared transition is 
smaller than those of other transitions (see Table \ref{tab:LDP}), and 
the S{\o}rensen-M{\o}lmer criterion
$\eta^2(\langle n\rangle+1)\ll1$
yields the condition $\langle n\rangle\ll 10^4$ that is easily fulfilled in practice.
In fact, a mean phonon number $\langle n\rangle<1$ can be reached: This
is shown in 
the middle part of Fig.~\ref{fig:ApAmin1}, where $\langle n\rangle$
is plotted vs the detuning of the 650-nm light. 
The lowest vibrational excitation was found by randomly searching 
the 4D parameter space for the lowest achievable 
$\langle n\rangle$. Then, three laser parameters were kept constant, while
the red-laser detuning was varied, in order to generate the graph.
The minimal temperature is achieved around 
$\Delta_{650}\approx 6 \times 2\pi$ MHz. Note that the lowest value  
 $\langle n\rangle=0.9$ indicated by the vertical dotted line
does not match a particularly large 
cooling rate, as is shown in the lower part of Fig.~\ref{fig:ApAmin1}.
Looking at the expected fluorescence signal (upper part of the
Figure), which corresponds
to the population of the P$_{1/2}$ level, we note that the optimum detuning
is on the red side of the dark resonance at 
$\Delta_{650}=4.9 \times 2\pi$ MHz, 
as is expected for Raman cooling.
The dark resonance is power broadened
by the high intensity at 650 nm. 
On the other hand, we find low temperatures only for low intensity
of the 493-nm light. This fact indicates that scattering of the blue light is the
dominant heating mechanism. We
do not claim to have found the absolute minimum of 
achievable temperature: 
Other sets of laser parameters may yield  similarly low temperatures, 
and
also the applied magnetic field to some degree influences  
the achievable temperature.

We have restricted the parameter search to the range of 
negative detuning. 
Even lower energies may be reached with both lasers detuned to the blue
side of the atomic resonance and the parameters adjusted to
yield so-called electromagnetically induced transparency cooling  
\cite{Mor00}. Here, the dynamic Stark contribution 
to the nonlinear susceptibility dominates the interaction of ions and
light \cite{Han70}.
However, Doppler precooling is still necessary and the
lasers must either be switched from negative to positive detuning, or else
additional light sources are 
required. Moreover, the intensities of the lasers must be precisely
controlled, since for this type of cooling to work, the dynamic 
Stark shift should 
match the vibrational frequency particularly well. However, cooling
mechanisms are desirable, that are more robust against variations of
intensity and detuning of the laser light, and thus are easier to
implement experimentally. 

\begin{figure}[htbp]
  \begin{center}
\epsffile{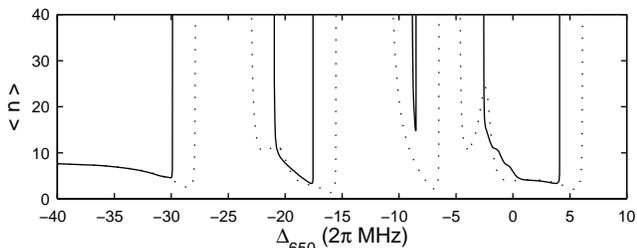}
    \caption{Mean number of vibrational quanta for fixed laser
      parameters (dotted) and averaged over nine combinations of
      both detunings (see text) (solid).
Laser parameters: $\Omega_{650}=59.5\times2\pi$ MHz, 
$\Omega_{493}=3.8\times2\pi$ MHz,
$\Delta_{493}=-10.9\times2\pi$ MHz. Other parameters as in Sec.~\ref{sec:Exp} }
    \label{fig:ApAmin_bad}
  \end{center}
\end{figure}

The robustness of the cooling is affected by the sharp boundaries
between parameter regimes of very low and high temperature, because in
these regions any drift of the lasers' frequencies may cause drastic
deterioration of the cooling. In order to find ``safe'' parameters we
repeated the search, incorporating a tentative drift in the following
way: For a given point in the $(\Delta_{493},\Delta_{650})$ parameter
plane, $\langle n\rangle$ was calculated at this point and at eight
additional points, where one or both of the detunings differ by
$\pm 1\times 2\pi$ MHz, and then averaged over the 
$\langle n\rangle$ values at all nine points.
The best robust parameters found differ from the ones used to
calculate Fig.~\ref{fig:ApAmin1} and the achievable ion temperature is
not as low as in the ideal situation, but one still finds $\langle
n\rangle=3.3$. The result is
shown in Fig.~\ref{fig:ApAmin_bad}, where
the mean phonon number is shown as a function of 
$\Delta_{650}$ (solid line). The dotted line shows, for comparison,
the vibrational excitation that is
achieved with the same parameters, when the averaging 
over nine points is {\em not} carried out. 
The regions of safe cooling are narrower, since detuning
 close to a ``boundary" between cooling and heating is penalized. 
It seems not advisable to use a rather narrow
region of cooling (as, for example, close to 
$\Delta_{650}\approx -7 \times 2\pi$ MHz), because spurious laser drift might
easily impose unexpected heating. However, some regions
[e.g. $\Delta_{650}\approx (-22$ to $-17) \times 2\pi$ MHz and
$\Delta_{650}\approx (-5$ to $ 5) \times 2\pi$ MHz] are broad enough to 
allow for some laser drift with satisfactory cooling.

\section{Conclusion}
\label{sec:Conc}
We have observed the effect of laser cooling and heating on the collective
motion of two trapped Ba$^+$ ions by monitoring the macroscopic
motional states of the ions.
These states indicate the vibrational excitation, or temperature of the ions.
In this way, Raman heating is demonstrated that 
dominates on the blue side of the dark resonances
(i.e.~$\Delta_{650}>\Delta_{493}+\Delta_{\text{Zee}}$).

The observed phase transitions are a remarkable
macroscopic manifestation of dark states, i.e. of the coherent
quantum superposition of atomic levels.
Raman cooling on the red side of the dark resonance is implied, though
it is not directly observed with this method, since no
macroscopic change of the motional state shows up.  

The good agreement of the calculated excitation 
with the experimental observations shows that
the  outlined theory describes well the laser cooling of the
collective motion of the ions. The predicted cooling rates much exceed
spurious heating rates of the vibrational motion, 
for example, from fluctuations of the 
trapping electric field. Thus, such heating is negligible as long as the 
cooling light is on.

The model comes with no free
parameters; however, it is applicable only if the
ions are localized near their potential wells. Nevertheless, this is
not a strong restriction since this regime is supposed to be
 used in any experiment implementing conditional quantum dynamics.
This allows us to identify parameter regimes 
that imply effective and robust laser cooling.
We find that the collective
motion can become cooled close to the ground state, if the dark
resonances are not much broadened by laser power or fluctuations. 
But even
when fluctuations in laser parameters are allowed for, low levels of
temperature emerge that 
are suitable for quantum information processing using
the scheme for conditional quantum dynamics suggested by 
S{\o}rensen and M{\o}lmer \cite{Sor99}. We note that the level scheme of Ba$^+$
has the same structure as, for example, Ca$^+$ \cite{Ste98}, a type of
ion often suggested, and the results obtained here 
 are applicable, with appropriate modifications, to the cooling of
these ions, too.

\bibliographystyle{prsty}


\end{multicols}

\end{document}